# Elastic Trapped States at Dislocation Defects in Scaled Coupling and Hofstadter Models

Yangkai Liu, [1] Cheng Lin, [1] Yuan Liu, [1] Jiao Shen, [1] Yifan Zhu,[1,2] Haiyan Fan, [1,2,†] Hui Zhang, [1,2,†]

1 Jiangsu Key Laboratory for Design and Manufacturing of Precision Medicine Equipment, School of Mechanical Engineering Southeast University, Nanjing, China.
2 Key Laboratory of Underwater Acoustic Signal Processing (Southeast University), Ministry of Education, Nanjing China.

Elastic topological dislocations provide a pathway for trapping elastic wave energy at internal defects, rather than being confined solely to external boundaries or corners, which are typically associated with topological insulators (TIs). However, two practical constraints persist. First, highly confined dislocation states based on conventional Su-Schrieffer-Heeger (SSH) dimerization usually require a large coupling contrast and a correspondingly enlarged bandgap, which may be challenging to realize. Second, some Hamiltonians with richer topological physics often contain complex hopping terms, synthetic gauge fields or nonlocal couplings, which substantially increase the geometric complexity of experimental samples. Here, dislocation-induced trapped states are demonstrated in both a scaled coupling (SC) model and a Hofstadter model (HM) within an elastic platform. In the SC model, the trapped mode is treated as a higher localized state in the continuum rather than an in-gap mode in the SSH model. Consequently, the SC-induced dislocation can trap an enhanced mode without the requirement of an enlarged bandgap. For the HM, Householder tridiagonalization is used to map the original tight-binding Hamiltonian with complex hopping terms onto a tridiagonal matrix with only positive-real-valued nearest-neighbour (NN) hopping terms. Truncation at a weak-hopping position preserves the topological phenomena and allows a dislocation defect to be constructed from the shortened aperiodic chain. The results establish a practical route for designing highly localized modes without relying solely on bandgap enlargement or complex couplings, which advance the topological physics of elastic wave systems and promise enhanced possibilities for elastic functional devices.

## 1. Introduction

Topological defect states arise when a real-space lattice defect introduces a symmetry flux that is topologically matched to the bulk band topology, as dictated by the bulk-defect correspondence—an index theorem that relates the existence of bound states at defects to the bulk topological invariants of the host material [1–3]. In ordered media, topological defects are usually classified by the symmetry or order parameter that is disrupted at their cores. These include disclinations, which break rotational symmetry and are characterized by a Frank angles; dislocations, which break translational symmetry and are characterized by Burgers vectors; and vortices, which carry a phase winding and are characterized by charge [1,4–6]. This viewpoint has recently been extended from solid-state electronic systems to classical wave platforms, including photonic [7], microwave [8], acoustic [9] and elastic metamaterials [10], which serve as macroscopic platforms where defect cores can be readily fabricated and characterized via direct measurements. In particular, the Su-Schrieffer-Heeger (SSH) model serves as a paradigmatic platform that is widely employed for constructing and studying various topological defects. Its conceptual simplicity and readily tunable topological phases make it an ideal testbed for exploring defect physics [1,11,12]. For instance, a two-dimensional (2D) weak topological insulator (TI) has been constructed in a magnetomechanical metamaterial by stacking one-dimensional (1D) Su-Schrieffer-Heeger (SSH) chains, and a bound state was experimentally trapped at the dislocation core [13]. However, for dislocation states built from SSH chains, stronger confinement is usually obtained by increasing the dimerization ratio. Highly localized states are therefore tied to a large hopping contrast and an enlarged topological gap [14–18]. In elastic-wave implementations, such a requirement is difficult to satisfy because the coupling strengths are determined by beam widths, plate dimensions, resonator couplings, or electromechanical loading, making it challenging to achieve a sufficiently large coupling contrast [19–22]. Although topological optimization and other methods can be used to construct large bandgaps [23], they often lead to structural discontinuities, thereby introducing fabrication sensitivity and geometric complexity.

Moreover, tight-binding Hamiltonians with complex or longer-range hopping terms often appear as essential ingredients for expanding the accessible topological phase space [24] and regulating band structures [25], winding numbers [26,27], topological transitions [25,28], bound-state multiplicities [27] and localization behavior [25,26]. For example, lattice models of three-dimensional (3D) time reversal invariant TIs contain spin-orbit-coupled hopping matrices with imaginary components [29], the Wilson–Dirac models for strong TIs can be formulated in a tight-binding representation in which the interlayer hopping matrices are purely imaginary [30], and $Na_2IrO_3$ tight-binding models (TBM) use long-range hopping amplitudes to drive topological phase transitions [31]. In addition, 3D acoustic TIs and related higher-order platforms have shown that classical waves can reproduce rich topological boundary and defect physics [32–39], but they also illustrate the experimental burden associated with multilayer coupling [40–42], symmetry control [43–49] and multi-resonator unit cells [50]. The Hofstadter model (HM) serves as another paradigmatic platform that hosts Chern-type topological phases when a synthetic magnetic flux threads the lattice, with its band topology tunable by the flux density [51]. Recent work has realized dislocation-induced helical modes in a 3D acoustic TI by stacking Hofstadter-model layers with complex hoppings [39]. Nevertheless, the inherent complex couplings necessitate a transformation to real-valued couplings and a multilayer architecture, significantly complicating sample fabrication.

In this study, these two issues are addressed by a scaled coupling (SC) model and an improved Householder tridiagonalization in the Hofstadter Hamiltonian, respectively. On the one hand, a SC model is introduced as a gapless platform to induce an enhanced trapped mode. Unlike the SSH chain, whose hopping terms alternate between $t$ and $st$, the SC chain employs an exponentially graded hopping sequence, i.e., $t$, $st$, $s^2t$, $s^3t$. This spatially graded hopping profile compresses the edge-state envelope. By stacking the 1D SC chains and cutting some sites, the 2D lattice with a dislocation defect harbors the terminal mode of the original 1D SC chain and thereby produces a compact trapped state at the dislocation core [13]. On the other hand, an improved Householder tridiagonalization is used to transform the original tight-binding Hamiltonian with complex hopping terms of HM into a tridiagonal matrix containing positive-real-valued nearest-neighbour (NN) hopping terms only [52,53]. This Householder tridiagonal reduction shows that the topological properties and eigenvalues of the original model can be inherited by a deterministic aperiodic 1D chain. Truncating the chains at weak hopping positions does not affect the mapped topological state [54–56]. On this basis, a mechanically feasible aperiodic 1D chain is obtained from the original 2D Hofstadter model and subsequently used to create a zero-dimensional (0D) trapped state. The good agreement among TBM predictions, finite-element simulations, and experimental measurements confirms that highly energy localization can be obtained in both gapless SC-based dislocation defects and Hofstadter-type dislocation systems. The results establish a practical route for designing highly localized modes without relying solely on bandgap enlargement or complex couplings, thereby offering a viable route toward the implementation of compact, high-performance elastic-wave devices, including sensors, filters, resonators, and other signal-processing systems.

## 2. Trapped State at a Dislocation Defect in the Scaled-Coupling Model

### 2.1 Tight binding models of scaled-coupling-induced dislocation defect

We start from the TBM of the SC model, it corresponds to a finite 1D chain with N sites and real-valued NN hoppings, as depicted in Fig. 1(a). In the site basis $|n\rangle$, the SC Hamiltonian is

$$H_{\mathrm{SC}} = \sum_{n=1}^{N-1} \kappa_n \left(|n\rangle\langle n+1| + |n+1\rangle\langle n|\right), \quad (1)$$

where $\kappa_n = t \cdot s^{n-1}$ is the hopping amplitude between sites $n$ and $n+1$. Here $|n\rangle$ denotes the localized basis state at the $n_{th}$ site, N is the total number of sites in the finite chain, $t$ is the reference hopping amplitude, and $s$ is the dimensionless scaling factor. For the SC chain considered here, the hopping amplitudes follow the scaled sequence

$$\kappa_n = ts^{n-1}, t > 0, 0 < |s| < 1, \tag{2}$$

where $t$ is the reference hopping strength and $s$ is the dimensionless scaling factor. Since Eq. (1) contains no onsite terms and only couples NN sites, the SC chain has a bipartite structure. The odd and even sites are assigned to sublattices $A$ and $B$, respectively,

$$|A_m\rangle \equiv |2m-1\rangle, \; |B_m\rangle \equiv |2m\rangle. \tag{3}$$

The chiral operator is defined as

$$\Gamma = \sum_{m=1}^{N_A} |A_m\rangle\langle A_m| - \sum_{m=1}^{N_B} |B_m\rangle\langle B_m|, \tag{4}$$

where $N_A$ and $N_B$ denote the numbers of sites on sublattices $A$ and $B$, respectively. The chiral operator satisfies

$$\Gamma^2 = I. \tag{5}$$

The Hamiltonian only couples sites belonging to different sublattices,

$$\Gamma H_{\mathrm{SC}} \Gamma^{-1} = -H_{\mathrm{SC}}, \tag{6}$$

or equivalently,

$$\{\Gamma, H_{\mathrm{SC}}\} = \Gamma H_{\mathrm{SC}} + H_{\mathrm{SC}}\Gamma = 0. \tag{7}$$

Equation (7) is the chiral-symmetry condition. If $|\psi_E\rangle$ is an eigenstate of $H_{\mathrm{SC}}$ with eigenvalue $E$,

$$H_{\mathrm{SC}} |\psi_E\rangle = E |\psi_E\rangle, \tag{8}$$

then

$$H_{\mathrm{SC}}\Gamma |\psi_E\rangle = -E\Gamma |\psi_E\rangle. \tag{9}$$

Therefore, all nonzero eigenvalues appear in pairs $(E, -E)$, whereas an unpaired eigenstate can occur only at $E = 0$. In the sublattice basis (A, B), the Hamiltonian takes the block-off-diagonal form

$$H_{\mathrm{SC}} = \begin{pmatrix} 0 & Q \\ Q^\dagger & 0 \end{pmatrix}, \tag{10}$$

where $Q: \mathcal{H}_B \rightarrow \mathcal{H}_A$ is the hopping matrix from sublattice $B$ to sublattice $A$. We consider an odd-site SC chain,

$$N = 2M + 1, \tag{11}$$

where $M$ is a positive integer. The numbers of sites on the two sublattices are

$$N_A = M + 1, \; N_B = M. \tag{12}$$

The matrix $Q$ is therefore an $(M + 1) \times M$ matrix and can be written as

$$Q = \begin{pmatrix} \kappa_1 & 0 & 0 & \cdots & 0 \\ \kappa_2 & \kappa_3 & 0 & \cdots & 0 \\ 0 & \kappa_4 & \kappa_5 & \ddots & \vdots \\ \vdots & \ddots & \ddots & \ddots & 0 \\ 0 & \cdots & 0 & \kappa_{2M-2} & \kappa_{2M-1} \\ 0 & \cdots & 0 & 0 & \kappa_{2M} \end{pmatrix}. \tag{13}$$

A real-space chiral index for the finite SC chain is defined as

$$\nu_{\mathrm{SC}} = \dim\ker Q^\dagger - \dim\ker Q. \tag{14}$$

According to the rank-nullity theorem,

$$\dim\ker Q = N_B - \operatorname{rank} Q, \tag{15}$$

and

$$\dim\ker Q^\dagger = N_A - \operatorname{rank} Q. \tag{16}$$

Subtracting Eq. (15) from Eq. (16) gives

$$\nu_{\mathrm{SC}} = N_A - N_B. \tag{17}$$

For the odd-site SC chain,

$$\nu_{\mathrm{SC}} = (M + 1) - M = 1. \tag{18}$$

The nonzero integer $\nu_{\mathrm{SC}}$ characterizes the finite SC chain as nontrivial in the real-space chiral-index sense. This index does not require translational symmetry or a Bloch Hamiltonian and is therefore applicable to the spatially graded SC chain. To establish the relation between the real-space chiral index and the zero-energy eigenstate, a general zero-energy eigenstate is written as

$$| \psi_0 \rangle = \begin{pmatrix} a \\ b \end{pmatrix}, \tag{19}$$

where $a$ and $b$ denote the modal-amplitude vectors on sublattices $A$ and $B$, respectively. The zero-energy eigenvalue equation

$$H_{\mathrm{SC}} | \psi_0 \rangle = 0 \tag{20}$$

is equivalent to

$$Qb = 0, \; Q^{\dagger} a = 0. \tag{21}$$

The total number of zero-energy modes is therefore

$$N_0 = \dim \ker Q + \dim \ker Q^{\dagger}. \tag{22}$$

Combining Eqs. (14) and (22) gives

$$N_0 \geq | \dim \ker Q^{\dagger} - \dim \ker Q | = | \nu_{\mathrm{SC}} |. \tag{23}$$

Since $\nu_{\mathrm{SC}} = 1$,

$$N_0 \geq 1. \tag{24}$$

Thus, the nonzero real-space chiral index guarantees the existence of at least one zero-energy mode. For the connected SC chain with all hopping amplitudes nonzero, $Q$ has full column rank. To demonstrate this property, we consider the square $M \times M$ submatrix formed by the first $M$ rows of $Q$,

$$Q_M = \begin{pmatrix} \kappa_1 & 0 & \cdots & 0 \\ \kappa_2 & \kappa_3 & \ddots & \vdots \\ 0 & \kappa_4 & \ddots & 0 \\ \vdots & \ddots & \ddots & \kappa_{2M-1} \end{pmatrix}. \tag{25}$$

Its determinant is

$$\det Q_M = \prod_{m=1}^{M} \kappa_{2m-1}. \tag{26}$$

Since all hopping amplitudes are nonzero,

$$\det Q_M \neq 0, \tag{27}$$

which implies

$$\operatorname{rank} Q = M. \tag{28}$$

Consequently,

$$\dim \ker Q = 0, \; \dim \ker Q^{\dagger} = 1, \; N_0 = 1. \tag{29}$$

Therefore, the odd-site SC chain is nontrivial in the finite-chain real-space chiral-index sense and supports exactly one zero-energy mode when all NN hopping amplitudes are nonzero [57,58].

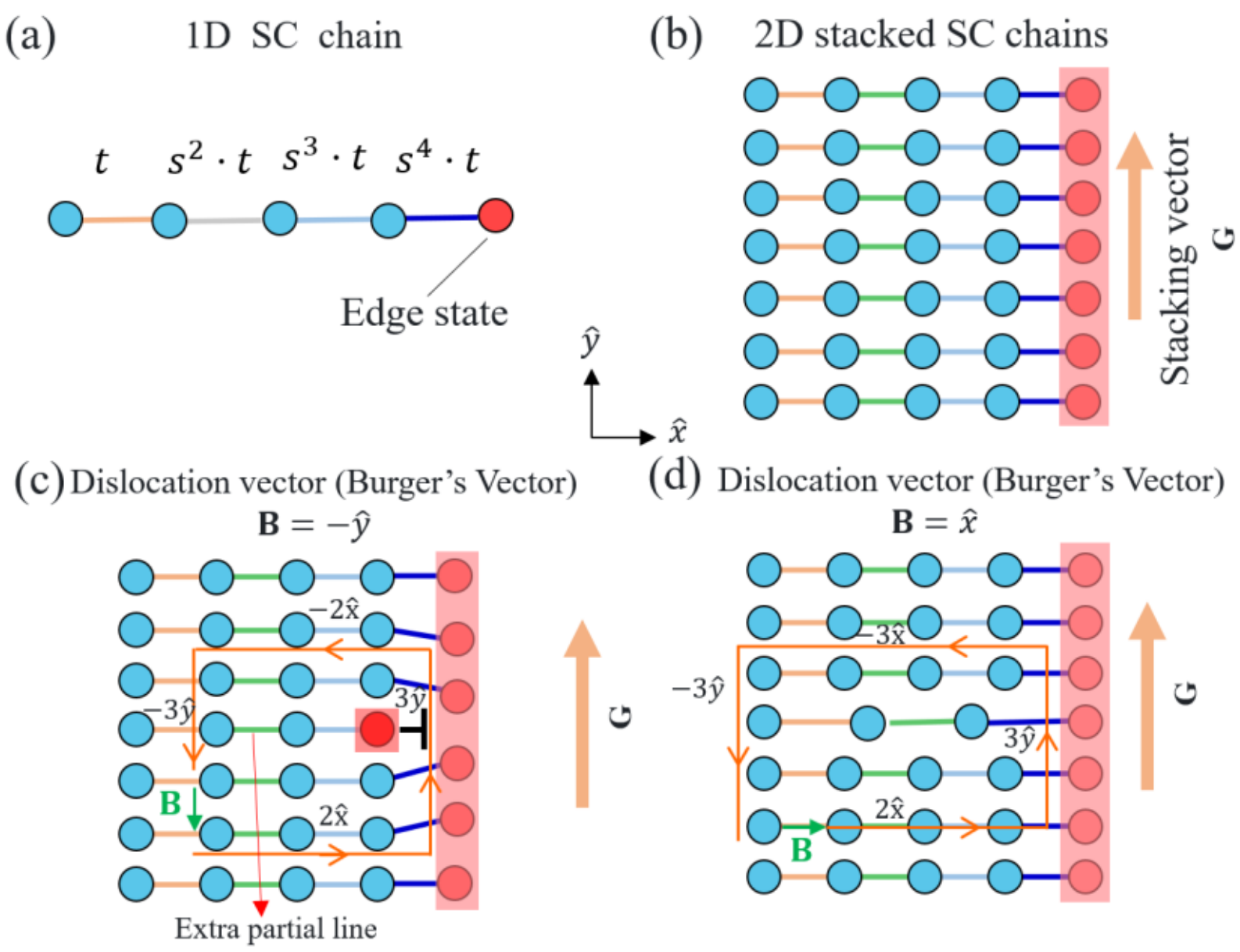

**Fig. 1.** Geometric construction of the dislocation in the SC model. (a) The 1D SC chain with edge state. (b) A 2D lattice formed by stacking the identical SC chains along the direction specified by the reciprocal stacking vector **G**. (c) A 2D lattice with dislocation defect and the Burgers vector $\mathbf{B} = -\hat{\mathbf{y}}$. Since B is parallel to G, there is a trapped state at the dislocation core. (d) A 2D lattice with dislocation defect and the Burgers vector $\mathbf{B} = \hat{\mathbf{x}}$. Since B is orthogonal to G, there is no corresponding trapped state.

In the schematic of Fig. 1(a), the SC chain is a spatially graded NN chain that runs along the $x$ direction with localized edge state at the right end. The 2D lattice in Fig. 1(b) is obtained by stacking the identical SC chains along the $y$ direction with lattice constant $a_y$. The reciprocal stacking vector can be written as

$$\mathbf{G} = (2\pi/a_y)\hat{\mathbf{v}}. \tag{30}$$

The vector **G** does not describe an additional hopping channel, rather, it identifies the direction along which the 1D SC chains are periodically repeated. This notation follows the stacked-chain dislocation construction used in Ref. [13], in which a reciprocal stacking vector is introduced to specify the direction of the layered 1D building blocks. A dislocation is introduced by creating a closure failure in the stacked lattice. For a Burgers circuit $C$ enclosing the defect core [as denoted by the rectangles in Figs. 1(c) and 1(d)], the Burgers vector is

$$\mathbf{B} = \oint_C d\mathbf{u}. \tag{31}$$

To illustrate the dislocation trapped state in the SC model, an edge dislocation, which is essentially a partial line [middle line in Fig. 1(c)], is introduced into the system. The right end of this partial line is the dislocation core. The number of stable dislocation modes is defined as $n = \mathbf{G} \cdot \mathbf{B}/2\pi$ modulo 2. For the example shown in Fig. 1(c), we have $\mathbf{G} = 2\pi\hat{y}$ (setting lattice constant $a_y = 1$) and $\mathbf{B} = -\hat{\mathbf{y}}$. Since **G** is parallel to $\hat{\mathbf{y}}$, this gives **B** parallel to **G**. In this case, the dislocation defect harbors the edge mode of the original 1D SC chain and hence produces a trapped state at the dislocation core. If **B** is orthogonal to **G**, however, no mode will be trapped at the dislocation defect [Fig. 1(d)]. Following the mechanism for constructing trapped states described above, the edge states and dislocation--induced trapped states based on the SSH and SC model were successfully reproduced by TBM calculations. In the theoretical calculations, the parameters $s$ and $t$ for both the SSH and SC models are set to 0.34 and 130, respectively [Figs. 2(a) and 2(b)]. The maximum edge--state intensity of the SSH chain is 0.445 [Fig. 2(c)], whereas that of the SC chain is much larger [Fig. 2(d)]. Hence, the SC model exhibits stronger localization than the SSH model under the same $s$ and $t$ parameters. The calculated eigen-energy distributions of the SSH--induced and S-C-induced dislocation models are plotted in Figs. 3(e) and 3(f), respectively. For the S-SH-induced dislocation model, the topological states, namely, dislocation states and edge states, appear within the bandgap [g-rey-shaded region in Fig. 3(e)]. In contrast, for th-e SC-induced dislocation model, both the trapped state (red dot) and the edge states (blue dots) are embedded inside the continuum of bulk states [Fig. 3(f)]. This indicates that t-he SC-induced trapped state is similar to a bound state in the continuum (BIC), implying that enhanced localization does not require a corresponding enlargement of the bandgap.

Besides, the calculated intensity at t-he SSH-induced dislocation core is 0.550 [Fig. 3(g)], which is substantially lower than the corresponding value of 0.832 for the SC-induced dislocation core [Fig. 3(h)], further confirming the superior localization capability of the SC-induced trapped state.

## 2.2 Numerical simulations of the scaled-coupling-induced dislocation in elastic lattices

The TBMs described above can be implemented in elastic wave via a plate-beam structure, in which the resonant plates correspond to the onsite potentials and the coupling beams represent the offsite hoppings [28]. The simulated results reveal that the edge mode is observed in the 1D SC chain [red dot in Fig. 3(a)]. The frequency response function (FRF) of the edge plate, shown by the red dotted curve in Fig. 3(b), exhibits a single pronounced peak at 1578 Hz. By contrast, the response of the selected bulk plate, shown by the blue dotted curve, contains several weaker resonances at other frequencies of bulk states [black dots in Fig. 3(a)]. The bulk-plate velocity amplitudes are substantially smaller than the edge-response peak. When identical 1D SC chains are stacked along the y direction and a dislocation defect is introduced, a trapped state emerges at the dislocation core [Fig. 3(c)]. In the 2D lattice, the response of the dislocation-core plate, represented by the red dotted curve in Fig. 3(d), displays a pronounced peak near 1570 Hz. The selected bulk plate, represented by the blue dotted curve, also exhibits a lower resonant peak at approximately 1570 Hz, together with resonances at several other frequencies. The coexistence of the defect resonance with bulk states at the same frequency confirms that the defect mode is similar to a BIC.

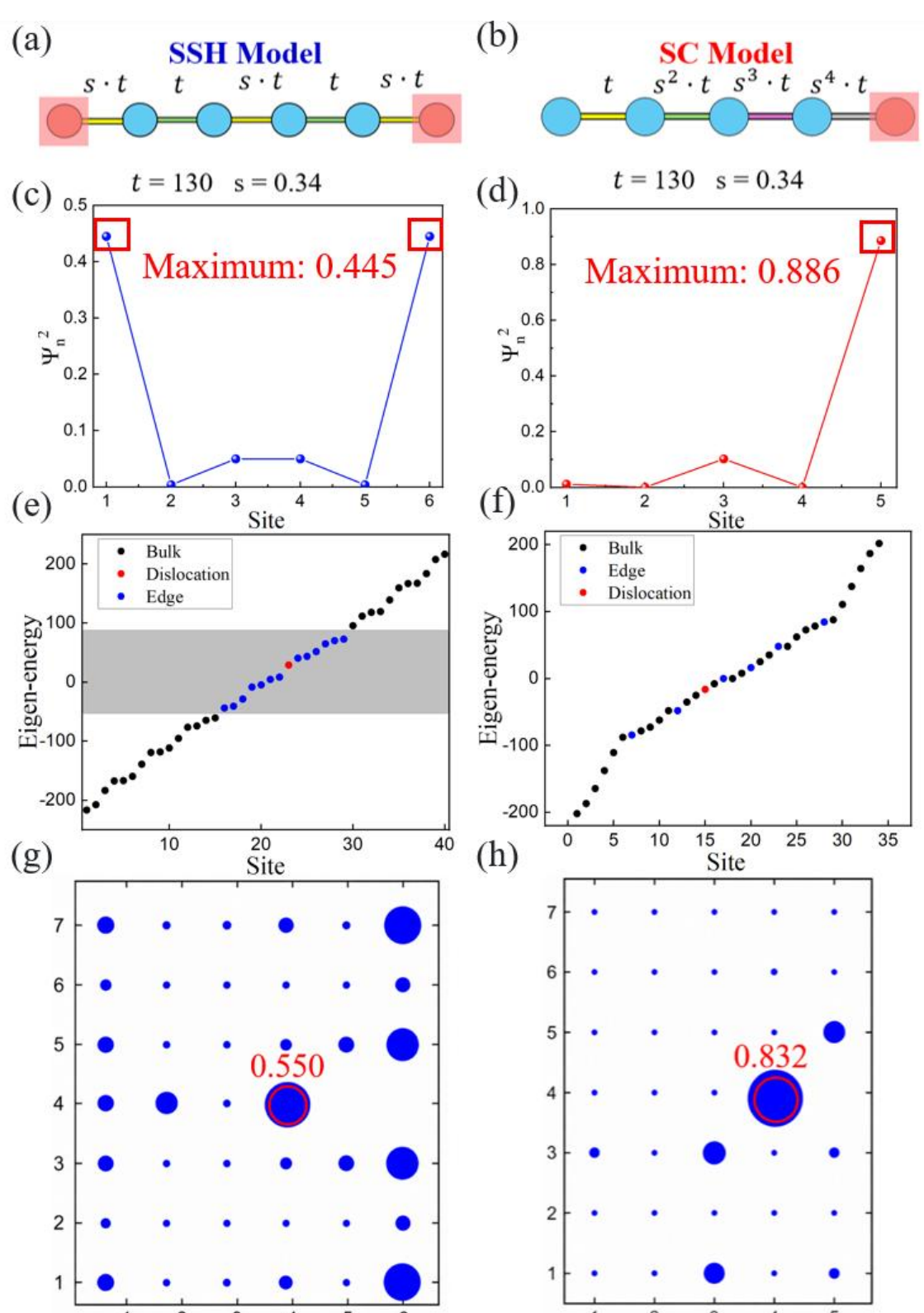

**Fig. 2.** TBM calculations of SSH and SC model. (a, b) Schematic of the SSH and SC models, respectively. (c, d) The intensity distributions of SSH and SC models, respectively. (e, f) The eigen-energy of the SSH-induced and SC-induced dislocation models. The shaded grey regions denote the bulk bandgap. Black dots, red dots and blue dots correspond to the bulk states, dislocation-induced trapped states and edge states, respectively. (g, h) Calculated intensity distributions of the dislocation states in the SSH-induced and SC-induced dislocation models, respectively.

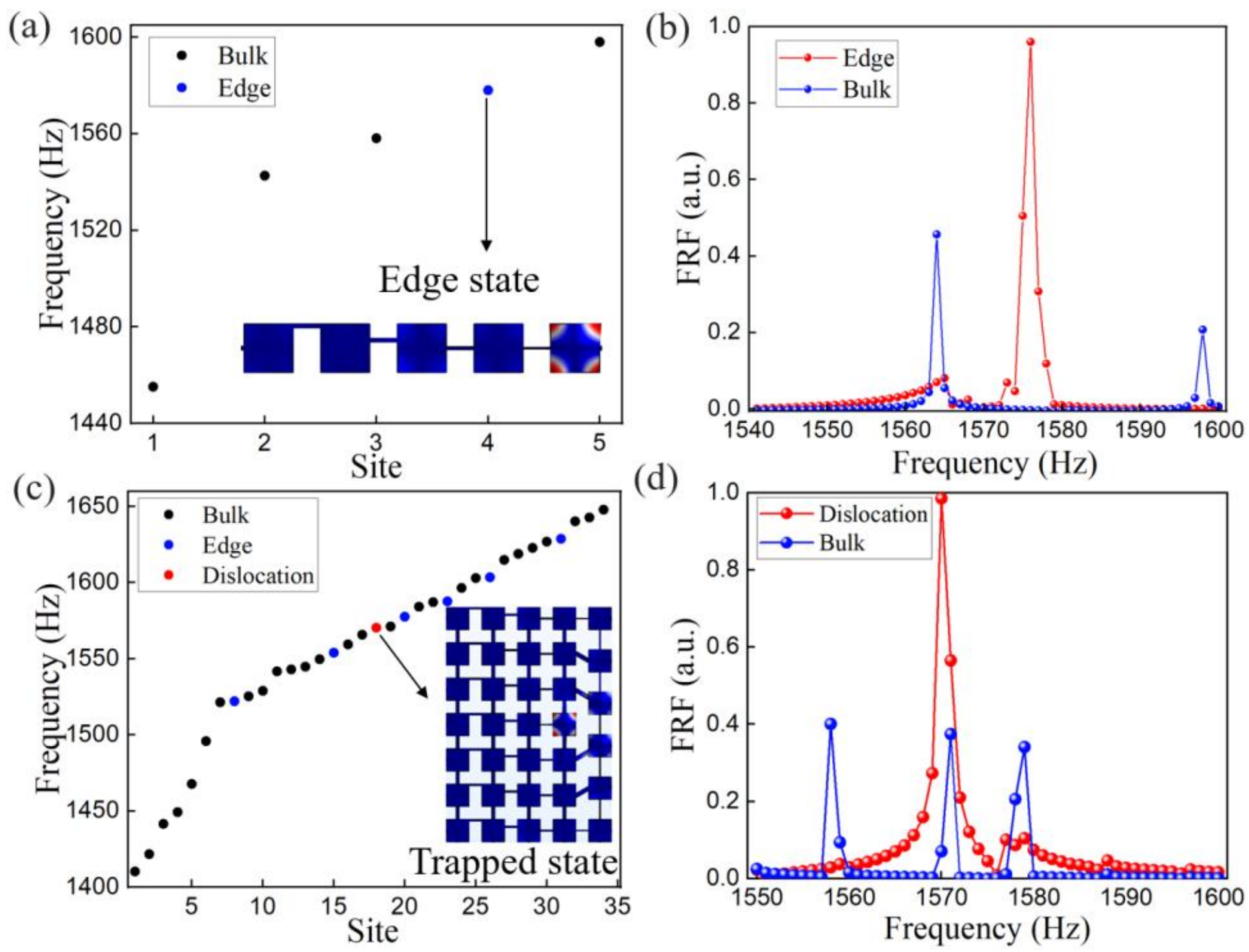

**Fig. 3.** Numerical results of SC-induced dislocation model. (a) Simulated eigenfrequencies for the 1D SC chain. The inset represents the out-of-plane displacement component of the edge state at 1578 Hz. (b) Simulated FRF spectrum of the bulk (blue dotted curve) and edge (red dotted curve) plates for the 1D SC chain. (c) Simulated eigenfrequencies for the 2D lattice with a dislocation defect. The inset denotes the out-of-plane displacement component of the trapped state at 1570 Hz. Black, red, and blue dots represent the bulk states, dislocation--induced trapped states, and edge states, respectively. (d) Simulated FRF spectrum of the bulk plate (blue dotted curve) and dislocation-core (red dotted curve) plate for the 2D lattice.

## 2.3 Experimental validation

The TBM predictions and the finite-element simulations are further validated experimentally using fabricated plate-beam samples [see Appendix A for detailed experimental set-up]. For the 1D finite SC chain, the measured mode shape of the edge plate at 1578 Hz is shown in Fig. 4(a), which matches well with numerical results in Fig. 3(a). The simulated and measured FRFs of the edge plate are compared in Fig. 4(b). The measured response, shown by the red dotted curve, exhibits a dominant resonance near 1578 Hz, consistent with the simulated response shown by the blue dotted curve. The minor secondary peaks and modest linewidth differences are attributed to material damping, fabrication tolerances, and nonideal boundary conditions. The measured site-resolved velocity-amplitude distribution at 1578 Hz, obtained by measuring all the plates, is presented in Fig. 4(c). It clearly shows a strong concentration of vibration on the edge plate, whereas the responses of the remaining plates are substantially weaker. The stacked dislocation specimen along with the measured mode shape of the defect plate at 1570 Hz is shown in Fig. 4(d). The measured FRF of the defect plate [red dotted curve in Fig. 4(e)] reproduces the simulated resonant peak near 1570 Hz [blue dotted curve in Fig. 4(e)]. The site-resolved velocity response profile at the peak frequency is obtained and displayed in Fig. 4(f), which shows the strongly concentrated vibration on the plate at dislocation core, whereas the surrounding plates exhibit substantially smaller responses [Fig. 4(f)]. These measurements verify the physical sequence predicted by the models: the gapless SC chain supports a strongly localized edge state, and geometric termination of an additional partial chain transfers this state to a trapped state located at the internal dislocation core.

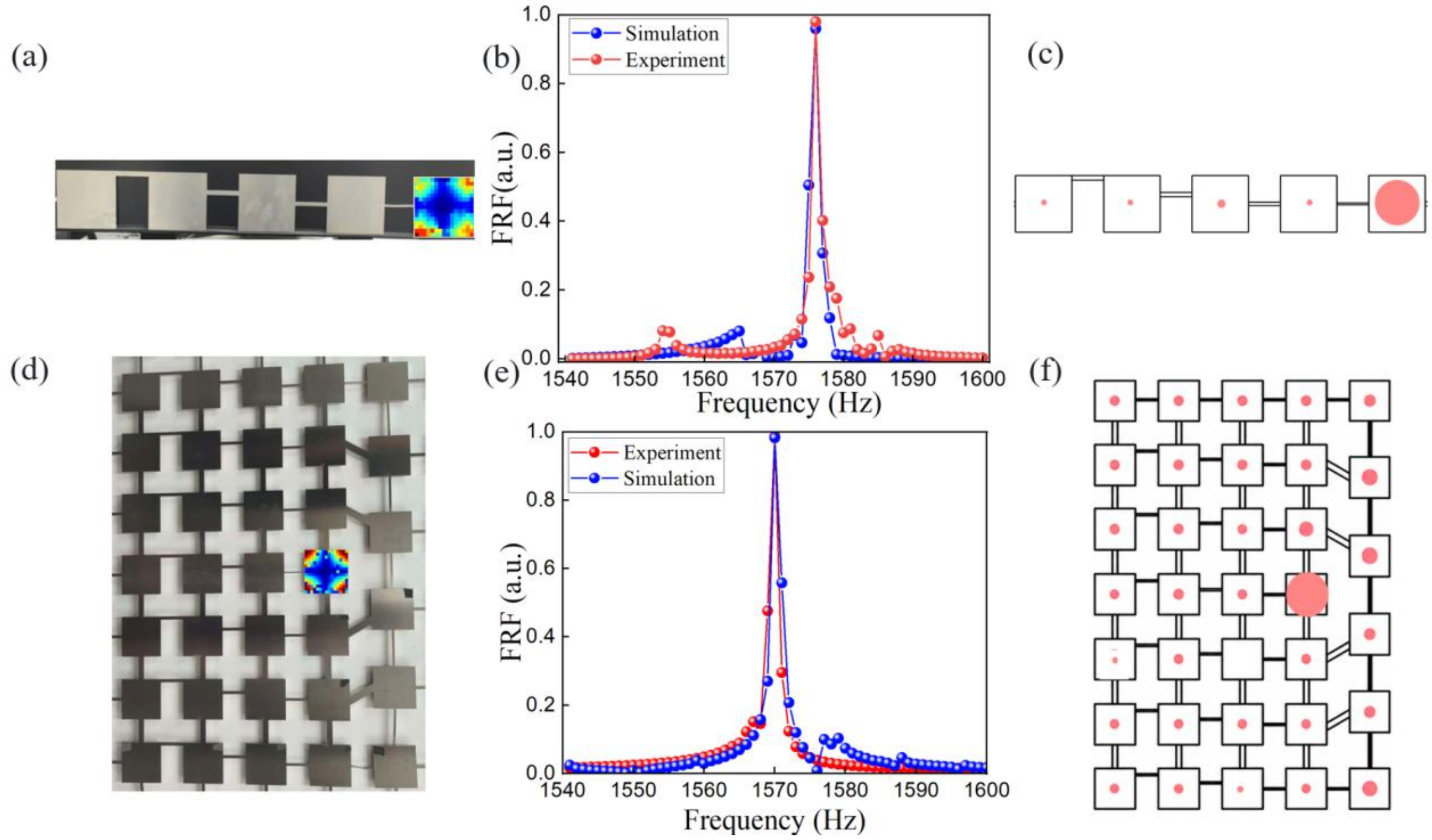

**Fig. 4.** Experimental validation of the finite SC chains and the SC-induced dislocation lattices. (a) Photograph of the 1D SC specimen. The inset shows the measured mode shape of edge plate. (b) The simulated (blue) and measured (red) FRF curves at the edge plate. (c) Velocity amplitude response measured on every single plate at 1578 Hz. The size of the pink circles is proportional to the vibration amplitude. (d) Photograph of the SC dislocation specimen along with mode shape at the dislocation core. (e) The simulated (blue) and measured (red) FRF curves at the defect plate. (f) Velocity amplitude response measured on every single plate at 1570 Hz.

## 3. Trapped State at a Dislocation Defect in Hofstadter Model

### 3.1 Tight binding models of Hofstadter-model-induced dislocation defect

The construction of the HM-induced dislocation is summarized in Fig. 5. The procedure consists of four steps: beginning with a 2D Hofstadter TI, applying Householder tridiagonalization, truncating the resulting chain at a weak hopping, and stacking the shortened local aperiodic chains to construct a dislocation defect. First, we consider a 2D strong TI formed by the HM carrying synthetic magnetic fluxes [Fig. 5(a)]. In the Landau gauge, the spin-resolved tight-binding Hamiltonian can be written as

$$H_{\mathrm{HM}} = -t_{\mathrm{H}} \sum_{m,n,\sigma=\pm 1} \left[c_{m+1,n,\sigma}^{\dagger} c_{m,n,\sigma} + e^{i\sigma m\phi} c_{m,n+1,\sigma}^{\dagger} c_{m,n,\sigma} + \mathrm{H.c.}\right], \tag{32}$$

where $c_{m,n,\sigma}^{\dagger}$ and $c_{m,n,\sigma}$ are the creation and annihilation operators at lattice site ($m$, $n$), respectively. The index $\sigma = \pm 1$ denotes the two pseudospin sectors, $t_{\mathrm{H}}$ is the NN hopping strength, and $\phi$ is the synthetic magnetic flux through each plaquette. In the present study, the magnetic flux is chosen as

$$\phi = \frac{\pi}{2}. \tag{33}$$

The two pseudospin sectors experience opposite magnetic fluxes and together form a time-reversal-invariant strong TI supporting topological boundary states. Second, Householder tridiagonalization is applied to simplify the Hofstadter Hamiltonian. At the $k$-th transformation step, the Householder matrix is defined as

$$P_k = I - 2\frac{v_k v_k^{\dagger}}{v_k^{\dagger} v_k}, \tag{34}$$

where $v_k$ is the corresponding Householder vector and $I$ is the identity matrix. The total transformation matrix is

$$Q = P_1 P_2 \cdots P_{N-2}. \tag{35}$$

The original Hofstadter Hamiltonian is then transformed according to

$$H_{\mathrm{tri}} = Q^{\dagger} H_{\mathrm{Hof}} Q. \tag{36}$$

After the transformation, the Hamiltonian matrix takes a tridiagonal form,

$$H_{\mathrm{tri}} = \begin{pmatrix} a_1 & b_1 & 0 & \cdots & 0 \\ b_1 & a_2 & b_2 & \ddots & \vdots \\ 0 & b_2 & a_3 & \ddots & 0 \\ \vdots & \ddots & \ddots & \ddots & b_{N-1} \\ 0 & \cdots & 0 & b_{N-1} & a_N \end{pmatrix}, \tag{37}$$

where $a_n$ denotes the onsite potentials [red squares in Fig. 5(b)] and $b_n$ denotes the NN hopping [black squares in Fig. 5(b)] between sites $n$ and $n+1$. Therefore, the original Hofstadter Hamiltonian containing complex hopping terms is directly transformed into a simplified form containing only positive-real-valued NN hopping terms. The transformed Hamiltonian preserves the eigenvalue spectrum and the topological states of the original HM [56]. Third, the tridiagonal Hamiltonian matrix generally contains several relatively weak hopping terms. By truncating the chain at one of these weak-hopping positions, a shortened Hamiltonian can be obtained. If the chain is cut at the $M$-th site, the retained Hamiltonian is written as

$$H_{\mathrm{retained}} = \sum_{n=1}^{M} a_n \mid n\rangle\langle n \mid + \sum_{n=1}^{M-1} b_n \left(\mid n\rangle\langle n+1 \mid + \mid n+1\rangle\langle n \mid\right). \tag{38}$$

The Hamiltonian $H_{\mathrm{retained}}$ represents a shortened aperiodic chain with spatially varying NN hoppings. Truncation at a sufficiently weak hopping can ease the implementation of experiments and inherit topological properties [red circle with a red dot in Fig. 5(c)] of the original Hamiltonian [55,56]. Finally, identical aperiodic chains obtained above are stacked to form a 2D lattice. Following the dislocation construction introduced in Section 2, part of one additional chain is removed so that the chain terminates inside the lattice [Fig. 5(d)]. In this way, a HM-induced dislocation lattice is successfully constructed. The inherited terminal state of the local aperiodic chain is transferred from the external boundary to the endpoint of the partial line. Consequently, a highly localized trapped state appears at the dislocation core, as illustrated in Fig. 5. The eigen-energy distributions of the transformed tridiagonal Hamiltonian matrix [red dots in Fig. 6(a)] shows good agreement with the distributions of the original matrix [black circles in Fig. 6(a)]. Figures 6(b) and 6(c) depict diagonal and off-diagonal values of the transformed Hamiltonian. The onsite potentials $a_n$ of the transformed matrix remain zero [Fig. 6(b)], while the hopping terms $b_n$ form an aperiodic sequence in Fig. 6(c). Cutting at the seventh hopping, which approaches zero, leads to a seven-site chain with the on-site potentials and off-site hoppings denoted in Figs. 6(d) and 6(e), respectively. The intensity distribution of the edge state of the shortened chain is represented in Fig. 6(e), demonstrating that the topological edge property of the original HM preserves after the reduction and truncation. To create an interior defect, the aperiodic chains are stacked and cut some sites of the extra line to form dislocation. Fig. 6(f) shows that the resulting trapped state is concentrated at the dislocation core. Hence, the Householder mapping preserves the original boundary dynamics, and the dislocation relocates that boundary response into the interior of the stacked lattice.

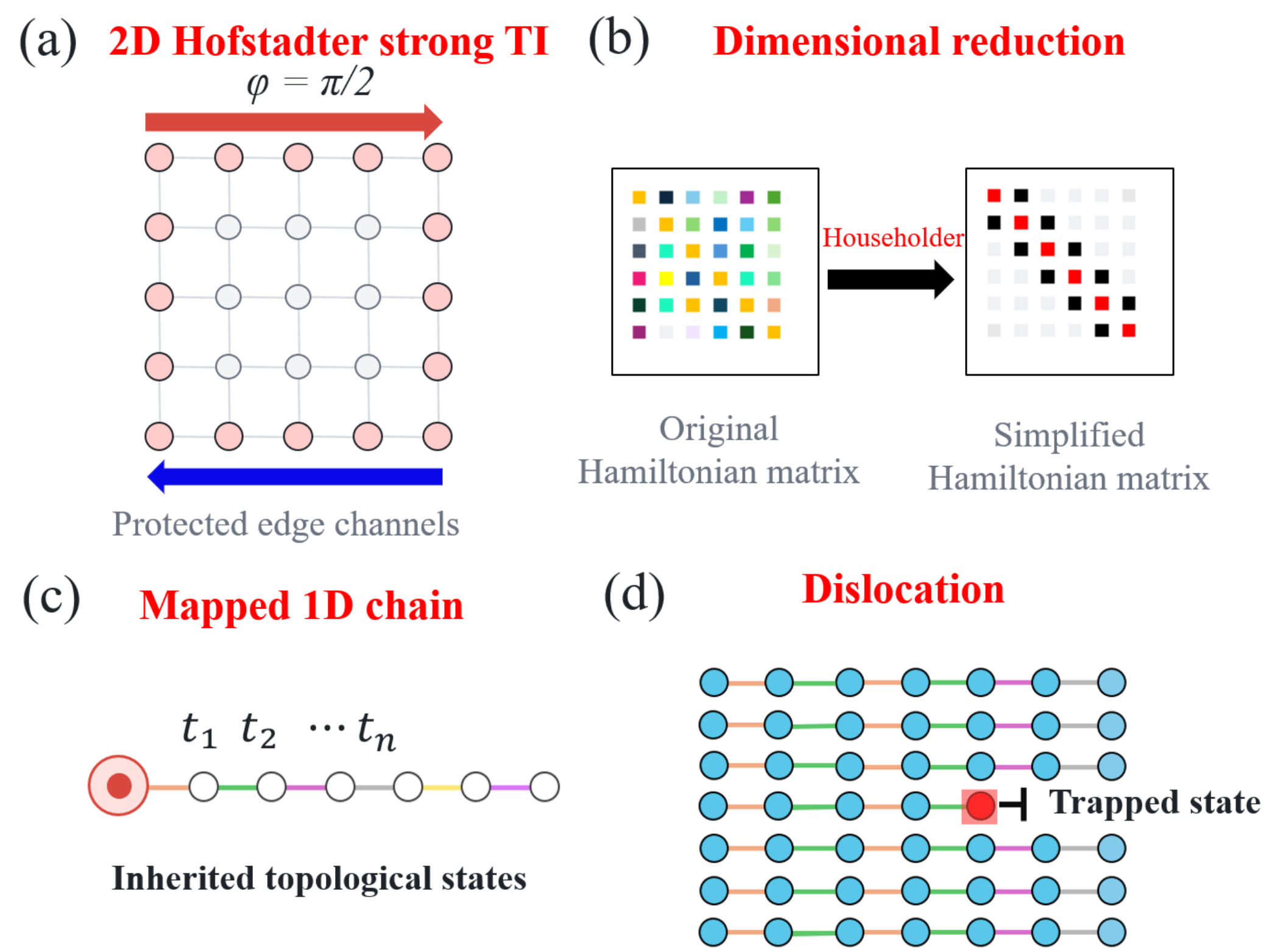

**Fig. 5.** Construction procedure of the HM-induced dislocation lattice. (a) A 2D strong TI is constructed from two time-reversed Hofstadter lattices with opposite synthetic magnetic fluxes $\phi = \pi/2$. (b) Householder tridiagonalization transforms the original Hofstadter Hamiltonian into a tridiagonal Hamiltonian. The black squares in the simplified matrix represent the positive-real-valued NN hopping terms. (c) The transformed chain is truncated at a weak-hopping position to a shortened aperiodic chain without changing the topological properties of the original HM. (d) Aperiodic chains are stacked and geometrically terminated, producing a trapped state localized at the dislocation core.

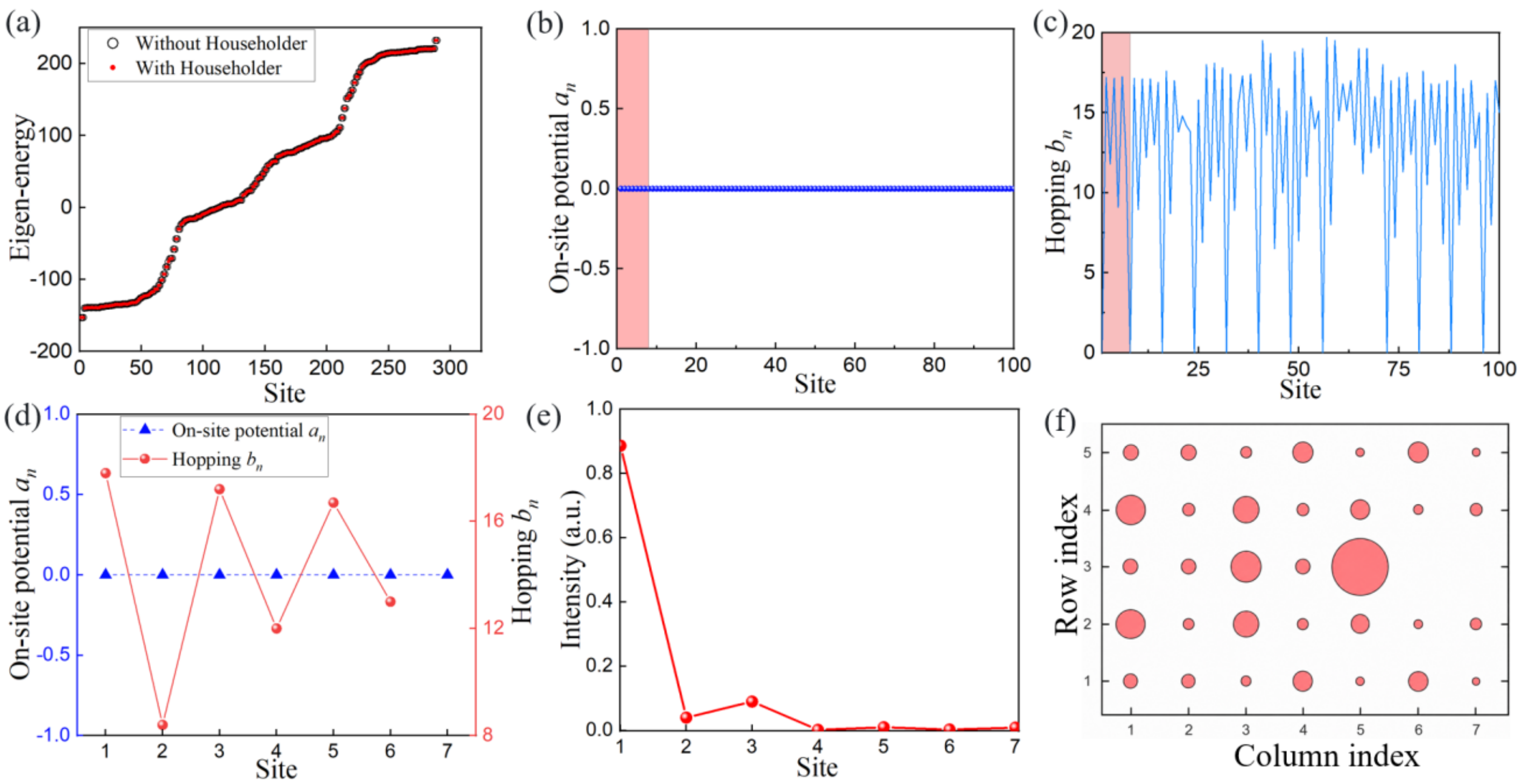

**Fig. 6.** TBM calculation of the Hofstadter-model-induced dislocation. (a) Eigenvalue spectra before (black circles) and after (red dots) Householder tridiagonalization. (b, c) On-site potentials $a_n$ and NN hoppings $b_n$ of the first 100 sites of the full tridiagonal Hamiltonian. The shaded pink region marks the retained segment after truncation. (d) On-site potentials $a_n$ (dashed blue line) and hoppings $b_n$ (solid red curve) of the truncated seven-

site local aperiodic chain. (e) Intensity distribution of its inherited edge-localized state. (f) The corresponding intensity distribution of the dislocation bound state based on local aperiodic chains.

## 3.2 Numerical simulations of Hofstadter-model-induced dislocation in elastic lattices

The transformed tight-binding parameters are mapped to a plate-beam model. As on-site potential $a_n$ are fixed at zero, identical square plates are used to represent these unchanged sites, while the fluctuated hopping $b_n$ are fulfilled by adjusting the geometrical parameters of the connecting beams [28]. The constructed plate-beam model of the shortened chain is shown in the inset of Fig. 7(a). We can clearly observe that the Householder tridiagonalization lift the requirements of multilayer experimental samples [14] to reproduce a synthetic flux. The simulated displacement field of the inherited edge states at 1558 Hz shows energy concentration on the anchor plate [first plate in Fig. 7(a)], corresponding to the sharp edge resonance [red dotted curve in Fig. 7(b)] in the FRF responses. The constructed dislocation structure is depicted in the inset of Fig. 7(c). The strong response at 1558 Hz of the 1D chain moves to the interior dislocation core of the 2D elastic lattice at 1558 Hz [Fig. 7(c)]. The FRF curve of the plate at the dislocation core [red dotted curve Fig. 7(d)] manifest a predominant peak at 1558 Hz [red dotted line in Fig. 7(d)]. These results confirm that the strongly localized dislocation-induced trapped state can still exist in the transformed HMs.

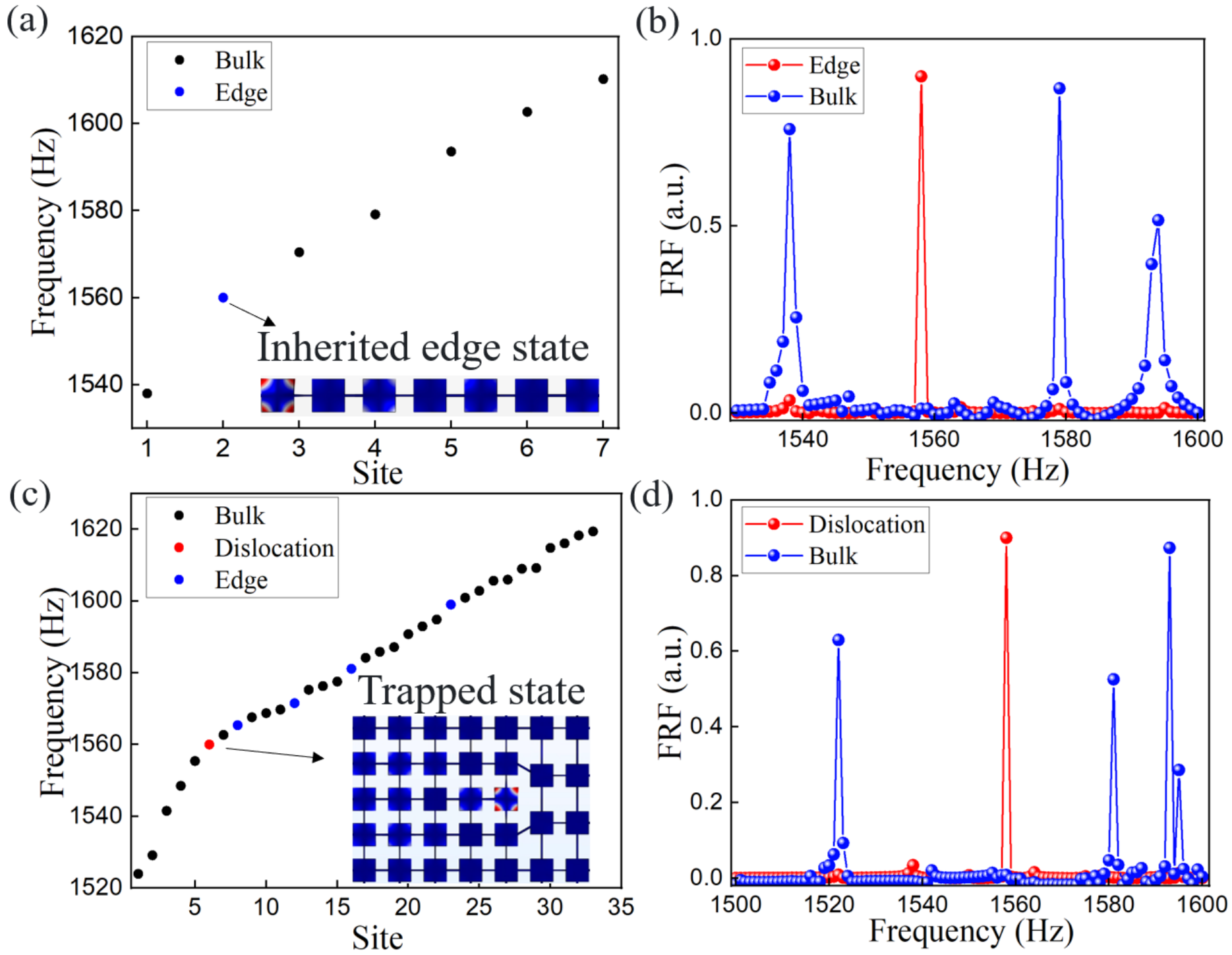

**Fig. 7.** Numerical results for the Hofstadter-derived elastic lattices. (a) Simulated eigenfrequencies of the 1D local aperiodic chain; the inset shows the out-of-plane displacement field of the edge state at 1558 Hz. (b) Simulated FRFs of the edge plate and a representative bulk plate in the 1D local aperiodic chain. (c) Simulated eigenfrequencies of the 2D dislocation-defect model; the inset shows the out-of-plane displacement field of the trapped state at 1558 Hz. The bulk states, dislocation--induced trapped states, and edge states correspond to black, red, and blue dots, respectively. (d) Simulated FRFs of the dislocation-core plate and a representative bulk plate.

## 3.3 Experimental validation

The local aperiodic chain and dislocation-defect specimens were characterized using the same apparatus and data-processing procedure as those used for the SC specimens. For the 1D local aperiodic chain, Fig. 8(a) shows the specimen and the measured mode shape of the edge plate at 1558 Hz, which agrees well with the numerical result in Fig. 7(a). The measured FRF, represented by the red dotted curve in Fig. 8(b), exhibits a narrow resonance near 1560 Hz and agrees closely with the simulated response shown by the blue dotted curve. The velocity-amplitude responses measured on all plates at 1558 Hz [Fig. 8(c)] demonstrate that the vibration is strongly localized on the anchor plate, whereas the responses of the remaining plates are substantially weaker. Figure 8(d) presents the dislocation-defect specimen and the measured mode shape at the dislocation core. The measured defect-plate FRF, shown by the red dotted curve in Fig. 8(e), exhibits a pronounced peak near 1580 Hz, in close agreement with the simulated resonance represented by the blue dotted curve. The plate-by-plate velocity-amplitude distribution at 1580 Hz [Fig. 8(f)] confirms that the response is highly localized at the dislocation core and strongly attenuated elsewhere. These experimental results demonstrate that a topological state of the original complex model can be inherited by a simple, locally aperiodic chain with positive-real-valued couplings and subsequently relocated to a dislocation core through stacking and geometric termination.

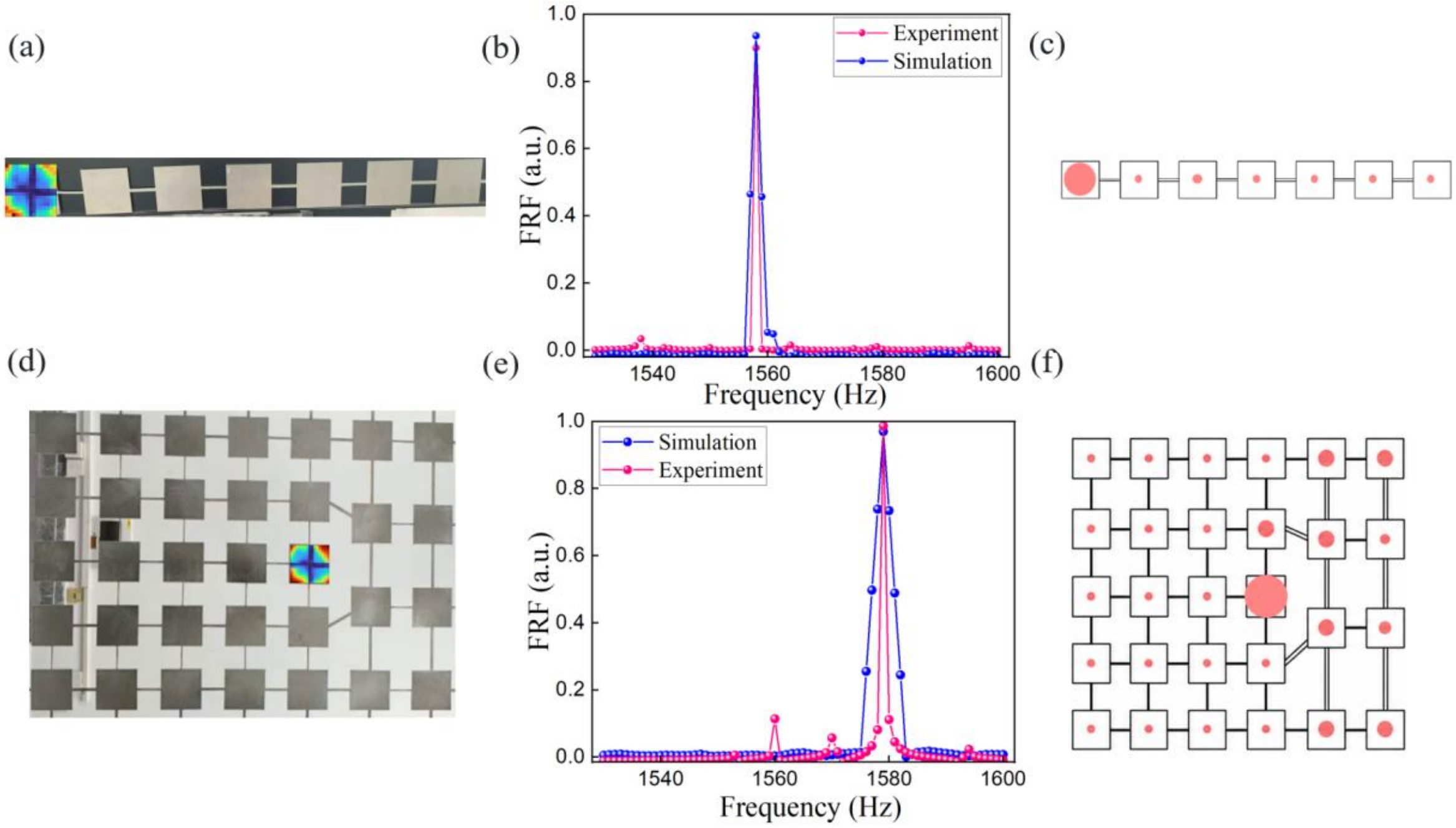

**Fig. 8.** Experimental validation of the local aperiodic chain and dislocation defect lattices based on the HM. (a) Photograph of the 1D local aperiodic chain specimen along with the measured mode shape of edge plate at 1558 Hz. (b) Simulated and measured FRF curves of the edge plate. (c) Velocity amplitude responses measured on all the plates at 1580 Hz. The size of the pink circles represents the vibration amplitude. (d) Photograph of the HM-induced dislocation specimen. The inset denotes the mode shape at the dislocation core. (e) Simulated and measured FRF curves at the defect plate. (f) Velocity amplitude response measured on all the plates at 1580 Hz.

## Conclusion

In this work, we demonstrated a strategy for trapping elastic-wave energy at engineered dislocation defects in plate–beam lattices. Dislocation defects were constructed in two distinct elastic systems—a gapless SC model and a HM model with complex hoppings—and the associated trapped modes were verified through tight-binding calculations, finite-element simulations, and experiments. By stacking SC chains and terminating a partial chain within the lattice, a strongly localized response is transferred from an external boundary to an internal dislocation core. Enhanced localization is therefore achieved in a gapless elastic lattice, relaxing the conventional SSH-based requirement that stronger confinement rely on increased dimerization contrast and the corresponding bandgap. We further established a practical implementation route for a HM with synthetic magnetic flux $\varphi = \pi/2$. Householder tridiagonalization maps the finite, complex 2D Hofstadter Hamiltonian onto a simplified Hamiltonian containing only onsite terms and positive-real-valued NN hoppings. After truncation at a weak coupling, the resulting reduced Hamiltonian can be implemented using a shorter chain while retaining the selected topological states of the original

Hofstadter model, thereby eliminating the need to fabricate complex samples. These results show that enhanced dislocation-induced trapped modes can be designed without relying exclusively on enlarged bandgaps or geometrically complex coupling architectures. The SC model provides a route to strong localization through a gapless chiral chain, whereas the Householder-reduced HM offers a practical and easier route for experimentally realizing the topological states stem from a complex Hamiltonian. Beyond providing new insights into defect-mediated elastic wave localization, these approaches establish a versatile design framework for next-generation elastic metamaterials, enabling compact and tunable platforms for energy confinement, wave manipulation, and advanced vibration-control functionalities.

## Acknowledgements

This work was supported by the National Natural Science Foundation of China (Grant No. 52501418), the Natural Science Foundation of Jiangsu Province (Grant No. BK20251266) and the National Natural Science Foundation of China (Grant No. 12574493)

## Appendix A. Experimental set-up

The sample was excited by an electromagnetic shaker (ZHENYAN TECH PME50) driven by a signal generator (Tektronix AFG31000) through a power amplifier (ZHENYAN TECH DPM200A). The excitation force $F(t)$ was measured by a force transducer (KETU YD-3200) and conditioned by a charge amplifier (KETU KT5852), while the out-of-plane vibration response $X(t)$ was acquired using a scanning laser vibrometer (HUAQIN OPTACOUS SLV-1550). A periodic chirp signal was used as the excitation signal to sweep the frequency range covering the simulated dislocation mode. The frequency-response function (FRF) was calculated as

$$H(\omega) = X(\omega)/F(\omega), \tag{A.1}$$

where $X(\omega)$ denotes the measured velocity response and $F(\omega)$ denotes the excitation force. The same signal path was used for both point-wise FRF measurements and full-field scanning of the mode profile. The signal path in measuring all the FRF spectra in this study is illustrated in Fig. A.1(a), and the picture of experimental set-up is shown in Fig. A.1(b).

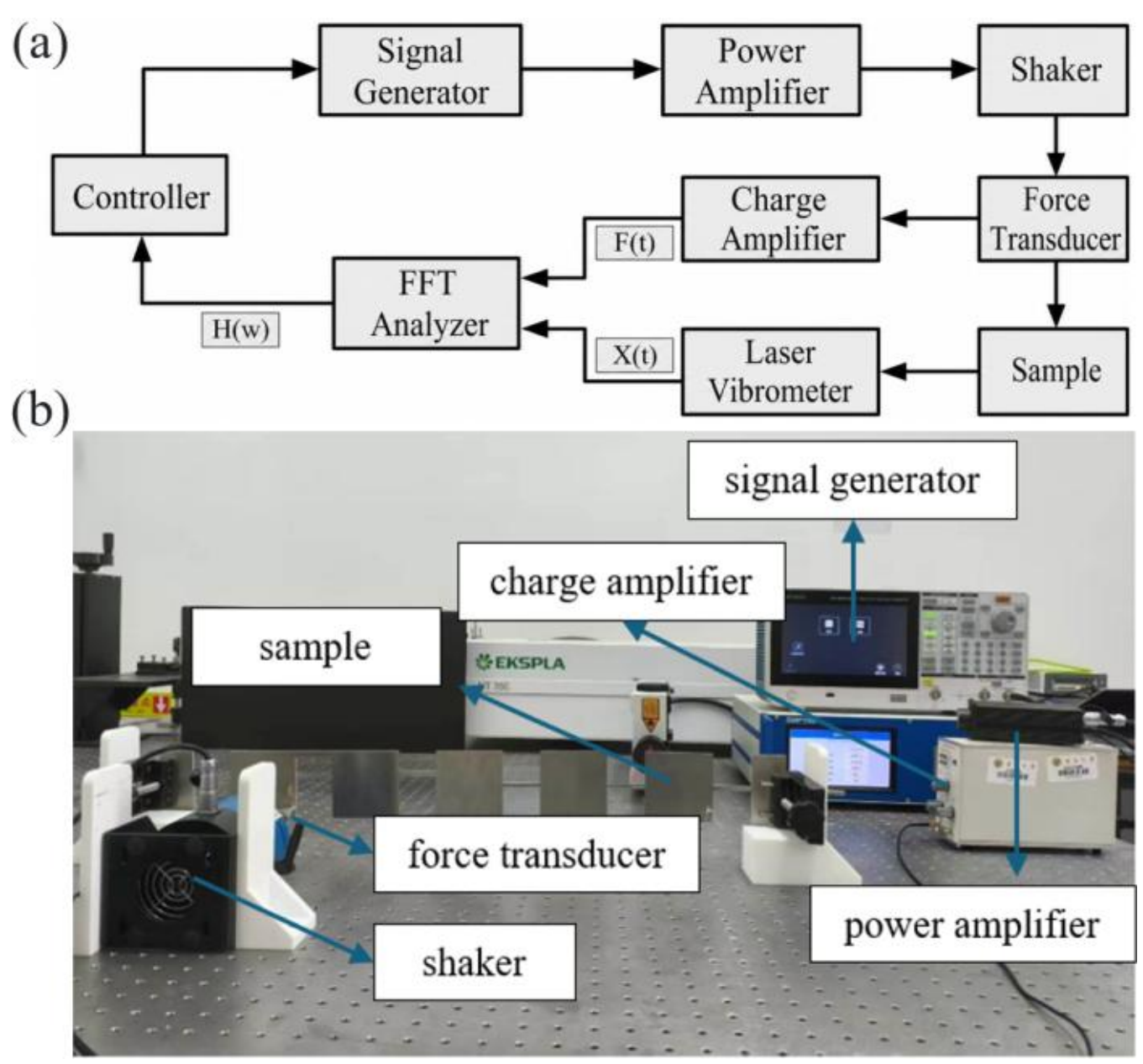

**Fig. A.1**. Schematics of experiments. (a) Signal path in measuring all the FRF spectra. (b) Experimental set-up.